\documentclass[graybox,natbib,nosecnum]{svmult}
\bibpunct{(}{)}{;}{a}{}{,} 

\usepackage{mathptmx}       
\usepackage{helvet}         
\usepackage{courier}        
\usepackage{type1cm}        

\usepackage{makeidx}         
\usepackage{graphicx}        
\usepackage{multicol}        
\usepackage[bottom]{footmisc}
\usepackage[normalem]{ulem}	
\usepackage{hyperref}  
\usepackage{graphicx}
\usepackage{amssymb}
\usepackage{amsmath}
\usepackage{verbatim}

\usepackage{soul}   

\newcommand{\hbindex}{}
\makeindex             

\begin{document}

\title*{Mapping Exoplanets}
\author{Nicolas B. Cowan and Yuka Fujii}
\institute{Nicolas B. Cowan \at McGill University, Montr\'eal, Québec, Canada\\ \email{nicolas.cowan@mcgill.ca}
\and Yuka Fujii \at National Astronomical Observatory of Japan, Mitaka, Tokyo, Japan\\ \email{yuka.fujii.ebihara@gmail.com}}
%
%
\maketitle

\abstract{The varied surfaces and atmospheres of planets make them interesting places to live, explore, and study from afar. Unfortunately, the great distance to exoplanets makes it impossible to resolve their disk with current or near-term technology.  It is still possible, however, to deduce spatial inhomogeneities in exoplanets provided that different regions are visible at different times---this can be due to rotation, orbital motion, and occultations by a star, planet, or moon.  Astronomers have so far constructed maps of thermal emission and albedo for short period giant exoplanets.  These maps constrain atmospheric dynamics and cloud patterns in exotic atmospheres. In the future, exo-cartography could yield surface maps of terrestrial planets, hinting at the geophysical and geochemical processes that shape them.}

\section{Introduction}
Astronomy often involves studying objects so distant that they remain unresolved point sources with even the largest telescopes. Exoplanets are particularly difficult to study because they are small compared to other astronomical objects. The diffraction limit dictates that resolving the disk of an Earth-sized planet at 10~pc requires a telescope---or array of telescopes---at least 25~km across at optical wavelengths, or 500~km across in the thermal infrared. Short of building a large interferometer to spatially \hbindex{resolve} the disks of exoplanets, we must rely on chance and astronomical trickery to map their atmospheres and surfaces. 

We expect planets to have spatial inhomogeneities: left to their own devices, atmospheres are subject to instabilities that produce spatially and temporally varying temperature, composition, and aerosols. Moreover, planets orbiting a star are further subject to asymmetric radiative forcing from their star, leading to diurnal (day--night) and seasonal (summer--winter) variations. Lastly, the surface of a planet may have a heterogeneous character due to, e.g., plate tectonics or volcanism.

Although part of the motivation to map exoplanets is simply to know what they look like, there are undoubtedly cases where understanding a planet requires understanding the diversity of its different regions. On a short-period planet, for instance, the hot dayside atmospheres may have atomic and partially ionized gas, while the nightside may be hundreds to thousands of degrees cooler, often shrouded in clouds and sometimes entirely airless. Interpreting the disk-integrated light of a heterogeneous planet with a one-dimensional model is misleading \citep[e.g.,][]{Feng_2016}. More importantly, the intrinsic climate of a planet is affected by its spatial inhomogeneities \citep{2023ApJ...957...20Z,2023ApJ...957...21Z,2023ApJ...957...22Z}. We would therefore like to map the atmospheres and surfaces of exoplanets. 

\section{Mapping Basics}
This chapter covers the analysis of planetary light, be it thermal radiation or reflected starlight.     
Planetary light can be separated from starlight temporally \citep{Charbonneau_2005,Deming_2005}, spatially \citep{Marois_2008}, or spectrally \citep{Brogi_2012}. Transmission mapping is also briefly described below.

\begin{figure}
\includegraphics[scale=0.23]{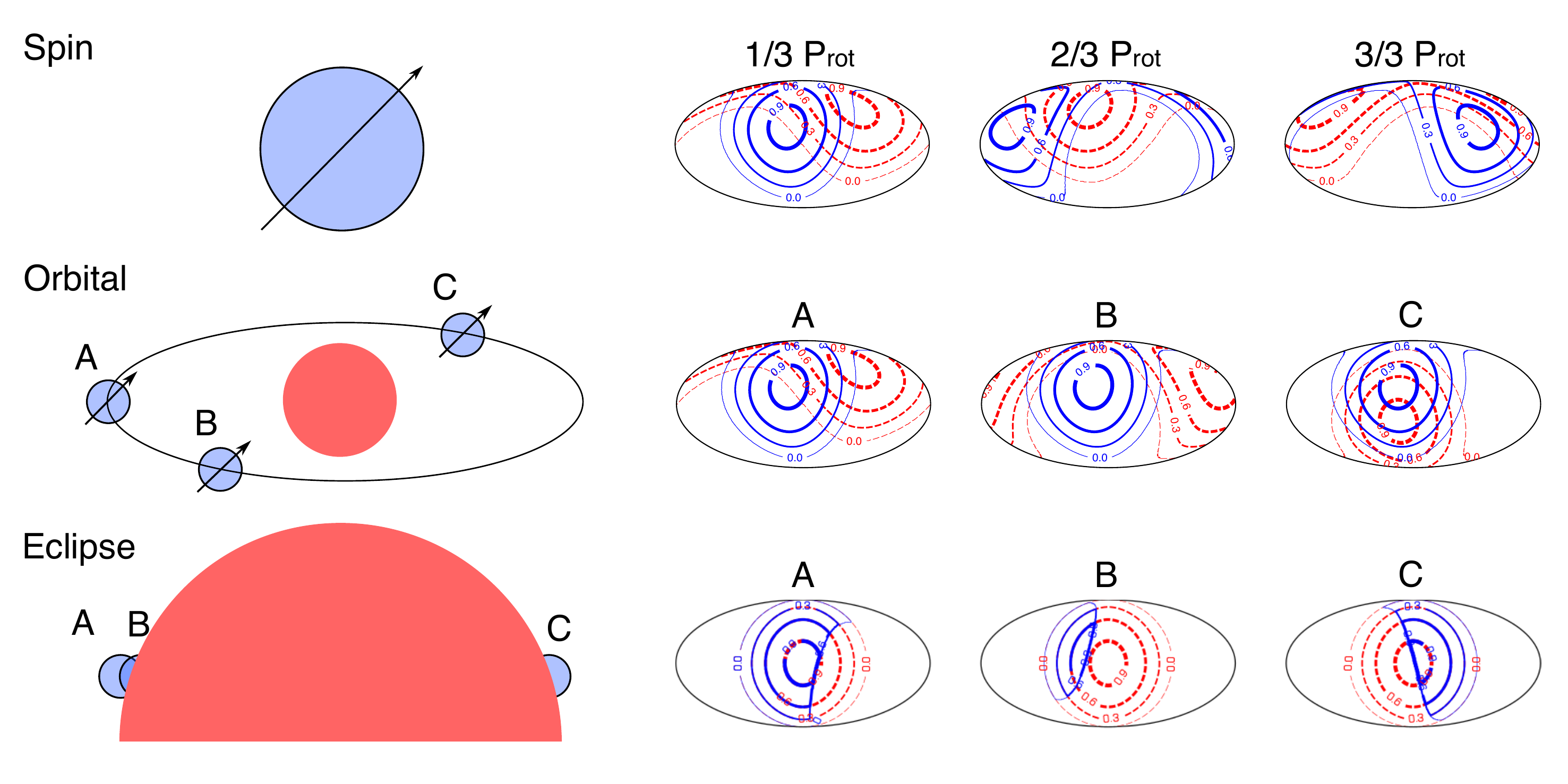}
\caption{The visibility (blue contours) and illumination (red contours) at three times for the different exoplanet mapping techniques: rotational mapping (top), orbital mapping (middle) and eclipse mapping (bottom). For the top two panels, we adopt an inclination of 60$^\circ$ and in the middle panel the planetary obliquity is 45$^\circ$.   For thermal emission, the convolution kernel is proportional to the visibility, while for diffuse reflected light the kernel is proportional to the product of visibility and illumination (see the Kernel section below). When the planet is unocculted (top two rows), the non-zero regions of the visibility and illumination are each hemispheres, making the non-zero portion of the reflection kernel a \emph{lune}. For the occultation geometry shown in the bottom row, we assume that the planet is slowly rotating so that essentially the same hemisphere is facing the observer before and after the occultation.}
\label{fig:kernel}
\end{figure}

It is possible to map the surface of an unresolved object as long as we don't always see the same parts of it---the resulting changes in brightness can be detected across astronomical distances. We are therefore able to tease out spatial information about a planet based on its time-variable brightness, provided we know something about the viewing geometry.  There are three ways in which our view of an exoplanet can change: rotation of the planet, changes in its position with respect to its star, and occultations by another object (Figure~\ref{fig:kernel}).

\hbindex{Rotational mapping} is possible because at any given time, we only see light from at most a hemisphere and as the planet spins on its own axis, different atmospheric and surface features rotate in and out of view. \hbindex{Orbital mapping} with reflected light is possible due to the changing illumination pattern as the planet orbits its star. \hbindex{Occultation mapping}, on the other hand, requires a second object to obscure part of the planet---the host star is the most likely culprit---hence changing our view and the system's overall flux. In practice, multiple sources of variability can be present in a single data set: phase variations are generally accompanied by rotation of the planet, and occultations are necessarily accompanied by orbital motion. 

\subsection{Transmission Mapping}
It is also possible to glean spatial information about an exoplanet's atmosphere when it transits in front of its host star: \hbindex{transmission mapping}, or transit mapping. 
Transmission mapping shares some concepts with emission and reflection mapping, so is briefly described here.  Transmission mapping relies on two geometrical facts: 1) only part of the \hbindex{planetary limb} is backlit during transit ingress and egress, and 2) the planetary limb probed in transmission moves on the planet during a transit.    

At the start of transit ingress, only the western (leading) limb of the planet is in front of the star and contributes to the transmission spectrum; likewise, the end of egress is only sensitive to the eastern (trailing) limb of the planet. Such transmission mapping has been proposed for detecting the difference in atmospheric scale height \citep{Dobbs-Dixon_2012}, winds \citep{Louden_2015}, and aerosols \citep{Kempton_2017,2019ApJ...887..170P} near the eastern and western terminators. Many codes have been developed to perform such transmission mapping  \citep{2021AJ....162..165E,2022ApJ...929...20M,2023MNRAS.519.5114G}.  The ingress-egress asymmetry is most marked for planets that are large compared to their star; the best spatial resolution could be obtained for a giant planet orbiting a white dwarf \citep[e.g.,][]{2020Natur.585..363V}. This phenomenon is conceptually similar to eclipse mapping, but is very sensitive to the limb-darkening of the star. Transmission mapping has already been performed with ground-based high resolution spectroscopy \citep[][]{2020Natur.580..597E}. Even in the middle of transit, high resolution data can differentiate between the blue-shifted eastern limb and and the red-shifted western limb \citep{2023MNRAS.522.5062B}, analogous to Doppler imaging.

Moreover, the planet will rotate during the transit, i.e., the planetary limb does not exactly coincide with the \hbindex{day--night terminator}. In fact, the limb only corresponds to the day--night terminator when the planet is in the center of the stellar disk, which never occurs for a non-zero impact parameter. At the start of transit the limb runs on the dayside of the eastern terminator and on the nightside of the western terminator, and vice versa at the end of transit. The angle between the limb and terminator is proportional to the angular size of the star as seen from the planet, and is therefore most important for ultra short period planets \citep{2020MNRAS.499.4605N}.  If a rocky planet has a negligible atmosphere, then topographical features may rotate in and out of view during the transit \citep{McTier_2018}. Transmission mapping of exoplanets is an exciting growth area, but the remainder of this review focuses on mapping exoplanets using light emitted or reflected by the planets themselves. 

\section{Mapping Formalism}
Exo-cartography is an \emph{inverse problem}, as opposed to the \emph{forward problem} of predicting the spectrum of a planet based on its surface and atmospheric properties.  The usual approach to an inverse problem is to solve the forward problem many times with varying parameters to see which ones best match the observations.

State-of-the-art forward modeling involves, at the very least, detailed radiative transfer calculations and hence is not useful for solving the inverse problem \citep{Robinson_2011}.
One of the necessities of tackling the inverse problem is therefore making judicious simplifications.  Adopting the notation of \cite{CFH_2013}, the exo-cartography forward problem is approximated as
\begin{equation} \label{convolution}
F(t) = \oint K(\theta, \phi, t) M(\theta, \phi) d\Omega, 
\end{equation}    
where $F(t)$ is the observed flux, or \hbindex{lightcurve}, $K(\theta, \phi, t)$ is the convolution \hbindex{kernel}, $M(\theta, \phi)$ is the top-of-atmosphere planet \hbindex{map}, $\theta$ and $\phi$ are planetary co-latitude and longitude, respectively, the differential solid angle is $d\Omega = \sin\theta d\theta d\phi$ for a spherical planet, and $\oint d\Omega$ is the integral over the entire planet. 
Although $F$ stands for ``flux'', the units of $F(t)$ depend on the situation and chosen normalization: planetary flux, planet/star flux ratio, reflectance, apparent albedo, etc.

Arguably the most important simplification is assuming a static map, ${\partial M}/{\partial t} = 0$, as it is generally intractable to map the surface of an unresolved planet when that map is changing. One can map a slowly varying planet, e.g., large-scale cloud patterns on Earth evolve slowly compared to the planet's rotation \citep{2020ApJ...900...48K,2022MNRAS.511..440T} and some eccentric planets may still be amenable to eclipse mapping, despite seasonal variations. Beyond these exceptions, one must adopt a parameterized time-variable map, $M(\theta, \phi, t)$. For example, phase curves of eccentric hot Jupiters 
 have been fitted with energy balance models \citep{Lewis_2013,2016ApJ...820L..33D,2022AJ....163...32D}. Finally, \cite{Cowan_2017} showed that for planets on circular, edge-on orbits one can infer a time-variable map based on the presence of odd harmonics; nonetheless, this is a far cry from mapping the changing surface of a planet.
 
Barring a changing map, time-variations in flux come in through the kernel. If there are no occultations, then the changing sub-observer and sub-stellar locations dictate the time-variability of the kernel and hence the time-variations in observed flux. 

\subsection{The Inverse Problem}
Inverse problems are often under-constrained, and exo-cartography is no exception: non-zero maps can produce flat lightcurves, a so-called \hbindex{nullspace}, and different maps sometimes produce identical non-trivial lightcurves. These pitfalls represent the loss of information: attempts to map distant objects suffer from formal degeneracies, even in the limit of noiseless observations.

If the orientation of a planet is not known \emph{a priori}, then the problem is even more challenging because the kernel is a function of one or more unknown parameters: spin frequency, spin orientation, etc. 
Nonetheless, it has been demonstrated in numerical experiments that one can extract a planet's spin (rotation rate, obliquity and its direction) from reflected lightcurves \citep[][]{Palle_2008, Oakley_2009, Kawahara_2010, Kawahara_2011, Fujii_2012, Schwartz_2016, Kawahara_2016,Farr_2018, Nakagawa_2020} or from thermal lightcurves \citep{Gaidos_Williams_2004, Gomez-Leal_2012, Cowan_Voigt_Abbot_2012, CFH_2013}. A similar formalism has been developed for imaging exoplanets with a solar gravitational lens \citep{2023MNRAS.525.5846T}. 

Exoplanets can be mapped in monochrome using single-band photometry, or using spectrally-resolved data to produce maps at each wavelength.   It is often more insightful, however, to assume that the lightcurves are correlated due to viewing geometry, molecular absorption, or surface spectra. With high spectral resolution, for example, the signal is in the time-variable mean molecular line shape. We point the interested reader to the excellent review of high resolution atmospheric characterization \citep{2018arXiv180604617B} and note that much work remains to be done in extracting spatial information from high resolution data. The inverse problem of mapping a planet using template spectra---let alone retrieving intrinsic surface colors and spectra---based on multi-band photometry or \hbindex{time-resolved spectroscopy} is beyond the scope of this review, but has been considered in the literature both in the context of reflected light \citep{Cowan_2009, Cowan_2011, Fujii_2010, Fujii_2011,  Cowan_Strait_2013, Fujii_2017,2020ApJ...894...58K} and thermal emission \citep{Kostov_Apai_2013, Buenzli_2014, Stevenson_2014, Feng_2016, Irwin_2020, Changeat_2020,2021ApJ...915...45C}.  

\section{Basis Maps and Basis Lightcurves}
The exo-cartography inverse problem boils down to fitting observed lightcurves.
One can use orthonormal \hbindex{basis maps}, e.g., spherical harmonics,
\begin{equation}
\frac{1}{4\pi}\oint{M_1(\theta,\phi)M_2(\theta,\phi)d\Omega} = \delta_{12},
\end{equation}
and repeatedly solve the forward problem in order to fit an observed lightcurve, or one can use orthonormal \hbindex{basis lightcurves}, e.g., a Fourier series or eigencurves \citep{Rauscher_2018},
\begin{equation}
\frac{1}{P}\int_0^P{L_1(t)L_2(t)dt} = \delta_{12},
\end{equation}
to directly fit the observed lightcurve, then reconstruct the associated map and its uncertainties in post-processing.  Sometimes, orthonormal basis maps produce orthonormal basis lightcurves and the problem is particularly tidy, as for rotational thermal mapping \citep{Cowan_Agol_2008,CFH_2013}. Usually, however, the adopted basis maps do not produce orthogonal lightcurves, and vice versa.

There are two broad classes of map parameterizations: global (e.g., spherical harmonics) and local (e.g., pixels). The optimal basis maps will depend on viewing geometry, expected map geometry, and the nature of the data \citep[cf.][]{Keating_2019,Beatty_2019}. In general, spherical harmonics are better for rotational mapping, smoothly varying maps, and full phase coverage. Pixels are preferable for eclipse mapping, sharp features, or partial phase coverage.  \cite{Apai_2017} introduced hybrid basis maps to fit rotational lightcurves of brown dwarfs: zonal bands harboring sinusoidal waves.

It is advisable to adopt a basis for which some maps are constrained and others are unambiguously in the nullspace, as it makes degeneracies easier to track down.  
As with many inverse problems, priors on the model parameters can help stabilize fits, especially when the data are noisy. For exo-cartography, this takes the form of enforcing map positivity \citep{Keating_2017,Farr_2018}, regularization to favour smoothness between pixels \citep[e.g.,][]{Knutson_2007a, Kawahara_2010, Kawahara_2011, Fujii_2012}, suppressing power in high-order modes with a sharp cutoff \citep[e.g.,][]{Cowan_Agol_2008, Majeau_2012} or a Gaussian process \citep{Farr_2018,2020ApJ...900...48K,2021JOSS....6.3071L}.

\subsection{Pixels and Slices}
\label{ss:pixel}
An intuitive approach is to pixelize the planetary surface. The most general pixelization scheme is 2-dimensional, for example, a latitude--longitude grid or Hierarchical Equal Area isoLatitude Pixelization (\href{http://healpix.sourceforge.net/}{HEALPix}), but more exotic schemes have been developed for occultation mapping \citep{Louden_2018}. 
\hbindex{Pixels} are useful if the kernel moves or changes its shape in more than one direction, e.g., eclipse mapping or rotational reflected light curves at multiple orbital phases. 

Denoting the intensity of the $i^{\rm th}$ pixel by $m_i$, we may express an arbitrary map as
\begin{equation}
M(\theta, \phi)  = \sum m_i M_i ( \theta, \phi ), 
\end{equation}
where the basis maps are
\begin{equation}
M_i ( \theta, \phi ) = \left\{
\begin{array}{ll}
1 & \;\;\;\;\;\; \mbox{if $\{ \theta, \phi \} \in$ $i^{\rm th}$ pixel}, \\
0 & \;\;\;\;\;\; \mbox{otherwise}. 
\end{array}
\right. 
\end{equation}

As small pixel is well approximated by a \hbindex{$\delta$-function} at its center.  Although $\delta$-functions are terrible for decomposing continuous maps, they have the advantage of being trivial to integrate \citep{Haggard_2018}:
\begin{equation} \label{delta_map}
F_{\theta\phi}(t) =  \oint K(\theta', \phi', t) \delta(\theta'-\theta, \phi'-\phi) d\Omega' = K(\theta, \phi, t).
\end{equation}    

For rotational mapping, latitudinal information is difficult to extract, so it is often reasonable to consider a map consisting of uniform \hbindex{slices}, like a beach ball. 
This situation includes rotational mapping in the thermal infrared \citep{Knutson_2007a} and rotational mapping of reflected light at a fixed orbital phase \citep{Cowan_2009}. 

\begin{figure}
\begin{center}
\includegraphics[trim={1.2cm 0 0 0},clip, scale=0.40]{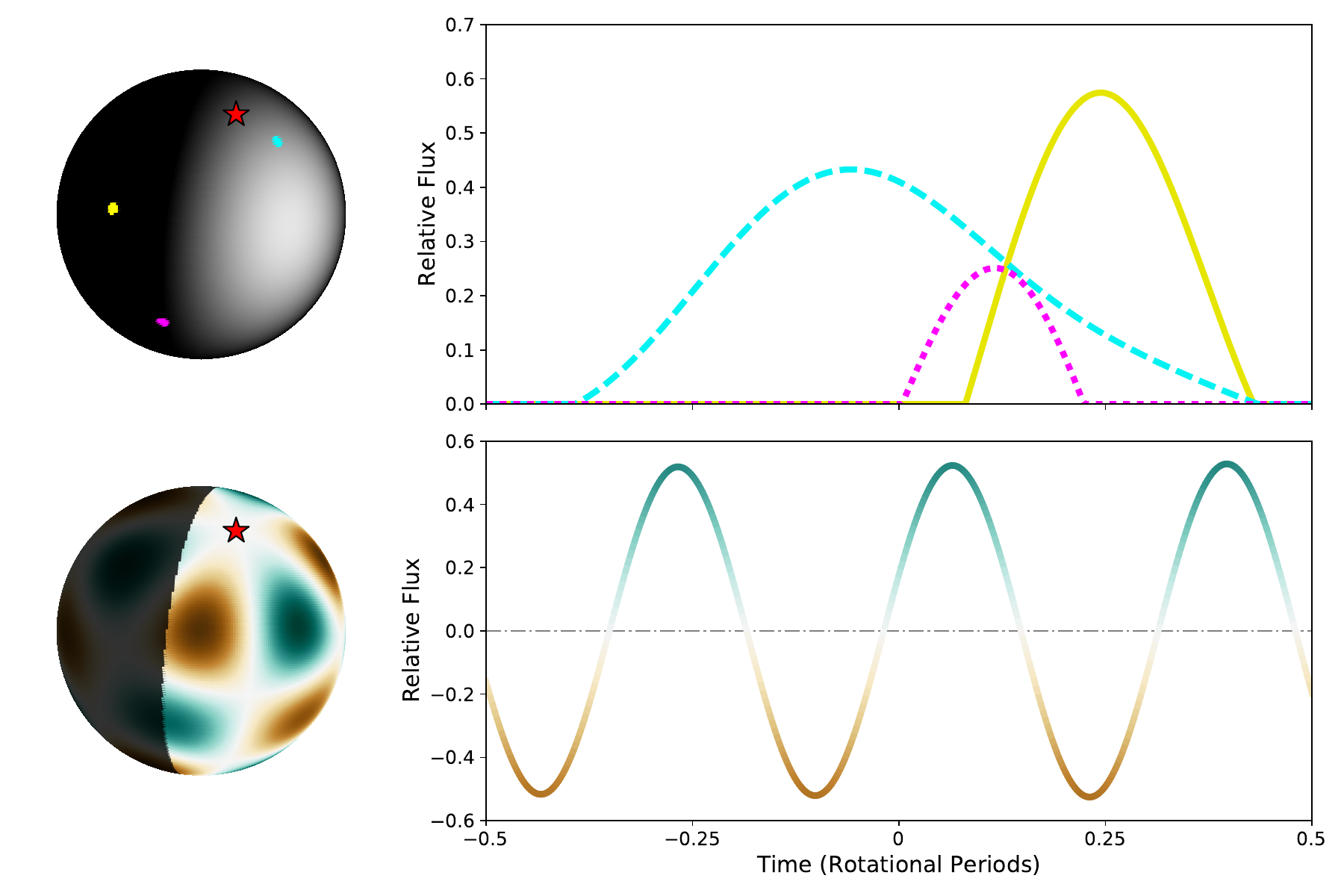}
\caption{Basis maps and their reflected lightcurves.  Examples of $\delta$-maps (top) and a spherical harmonic map (bottom). For the $\delta$-functions of the top panel, the resulting lightcurves are simply the value of the kernel at a particular location on the planet as a function of time, making them trivial to compute; small pixels would have approximately the same lightcurves. For the spherical harmonic map in the bottom panel, note that the resulting lightcurve is conveniently sinusoidal. The red star in each image denotes the north planetary pole, and the planet rotates to the east. Figure from J.C.\ Schwartz (private communication).}
\label{fig:basis_maps}
\end{center}
\end{figure}

\subsection{Spherical Harmonics}
A continuous albedo map, $M(\theta,\phi)$, may be decomposed into \hbindex{spherical harmonics}:
\begin{equation}\label{map_construction}
M(\theta, \phi) = \sum_{l=0}^{\infty} \hspace{0.2cm} \sum_{m=-l}^{l} C_{l}^m Y_{l}^{m}(\theta, \phi),
\end{equation}  
where the coefficients are
\begin{equation}\label{coefficients}
C_{l}^{m} = \frac{1}{4\pi}\oint M(\theta, \phi) Y_{l}^{m}(\theta, \phi) d\Omega.
\end{equation}

It is difficult to extract latitudinal information from purely rotational variations \citep[the little that can be squeezed out is described by][]{CFH_2013,Cowan_2017}. It is expedient in such cases to adopt sinusoidal basis maps \citep{Cowan_Agol_2008}, essentially spherical harmonics with $m=l$.

\subsection{The Kernel}
The kernel is the equivalent to the vertical contribution function in 1D atmospheric radiative transfer (Kawahara and Fujii have used ``\hbindex{weight}'' instead of ``kernel''). 
By definition, the kernel must move over the planet in order for exo-cartography to be possible. The shape of the kernel, however, is constant for rotational mapping and thermal phase mapping.  For such a static kernel, some maps are always in the nullspace, leading to severe degeneracy in the inverse problem \citep{CFH_2013}.   If the shape of the kernel changes with time, then maps need not remain in the nullspace, e.g., eclipse mapping \citep{2023AJ....166..176C}, or reflected-light mapping involving both rotation and orbital motion \citep{Haggard_2018}.  The degeneracy-busting power of a changing kernel is analogous to how a changing antenna pattern reduces the degeneracy in the location of a radio source on the sky.

\subsubsection{Thermal Emission}
For diffuse \hbindex{thermal emission}, the kernel is simply the normalized visibility:
\begin{equation}
K(\theta, \phi, t) = \frac{1}{\pi}V(\theta, \phi, t),
\end{equation}
where the \hbindex{visibility}, $V$, is unity at the sub-observer location (the center of the planetary disk as seen by the observer), drops as the cosine of the angle from the sub-observer location if limb-darkening can be neglected, and is zero on the far side of the planet or any part of the planet hidden from view by an occulting object.  
\cite{CFH_2013} derived analytic orbital/rotational lightcurves for spherical harmonic basis maps and \cite{Luger_2019a} did the same for occultations. 

Due to the unchanging kernel shape for rotational or orbital thermal mapping, a large fraction of spherical harmonics are in the nullspace, limiting the fidelity of retrieved maps. A common strategy is to ignore basis maps in the nullspace, setting their amplitude to zero in fits \citep[][]{Cowan_Agol_2008}. But this strategy can sometimes produce retrieved maps that appear to have regions of negative flux, an artifact of the band-limited deconvolution. The solution is to fit for more basis maps---including those in the nullspace: immeasurable fitted parameters may help the map remain everywhere positive \citep{Keating_2019}.  
Such hidden variables can be further constrained by a regularization scheme or Gaussian process. Indeed, the most general and robust approach to mapping is likely a high-order orthogonal map expansion---many spherical harmonics or pixels---embedded in a hierarchical model \citep{Farr_2018,2021JOSS....6.3071L}. If the prospect of fitting unseen and unknowable parameters by relying on priors is daunting, rest assured that you can often obtain a good fit and reasonable map by adopting an altogether different parameterization, such as slices or pixels \citep{Beatty_2019}.  

\subsubsection{Reflected Light}
For diffuse \hbindex{reflected light} (a.k.a. Lambertian reflection), the kernel is the product of visibility and illumination:
\begin{equation}
K(\theta, \phi, t) = \frac{1}{\pi}V(\theta, \phi, t) I(\theta, \phi, t),
\end{equation}
where the visibility is defined as above, and the \hbindex{illumination} $I$ is unity at the sub-stellar location (the center of the planetary disk as seen from the star), drops as the cosine of the angle from the sub-stellar location, and is zero on the night side of the planet or any part of the planet in the shadow of another object (e.g., an eclipsing planet or moon). 
Analytic reflected lightcurves have so far been derived for spherical harmonic maps at full phase \citep{Russell_1906}, for tidally-locked, edge-on geometry \citep{CFH_2013}, for the general case of an unocculted planet \citep{Haggard_2018}, and finally including occultations \citep{2022AJ....164....4L}. Extending this framework to polarized reflected light is beyond the scope of this review, but would undoubtedly yield useful constraints on cloud and surface properties \citep[for a review of exoplanet polarimetry, see][]{Wiktorowicz_2015}.

The Lambertian scattering approximation is suitable for mapping gibbous planets \citep{Fujii_2012}, but preferential back-scattering can be important at full phase, while forward scattering and specular reflection are important at crescent phases \citep{Robinson_2010}.  If \hbindex{specular reflection} is the dominant form of reflection, then the kernel is approximately a $\delta$-function at the location of the glint spot, making the forward problem analytically tractable \citep{Haggard_2018}.   
Specular reflection from surface liquid water is important as it is a direct indication of habitability \citep[][and chapter by Tyler Robinson]{Robinson_2014}. Combining Lambertian mapping at gibbous phases with glint mapping at crescent phases would provide strong evidence for surface liquids \citep{LustigYaeger_2018}: oceans appear dark at most phases, but bright when specularly reflecting grazing sunlight \citep[especially in polarized light;][]{Groot_2020}.

\subsubsection{Occultations}
\begin{figure}
\begin{center}
\includegraphics[scale=0.36]{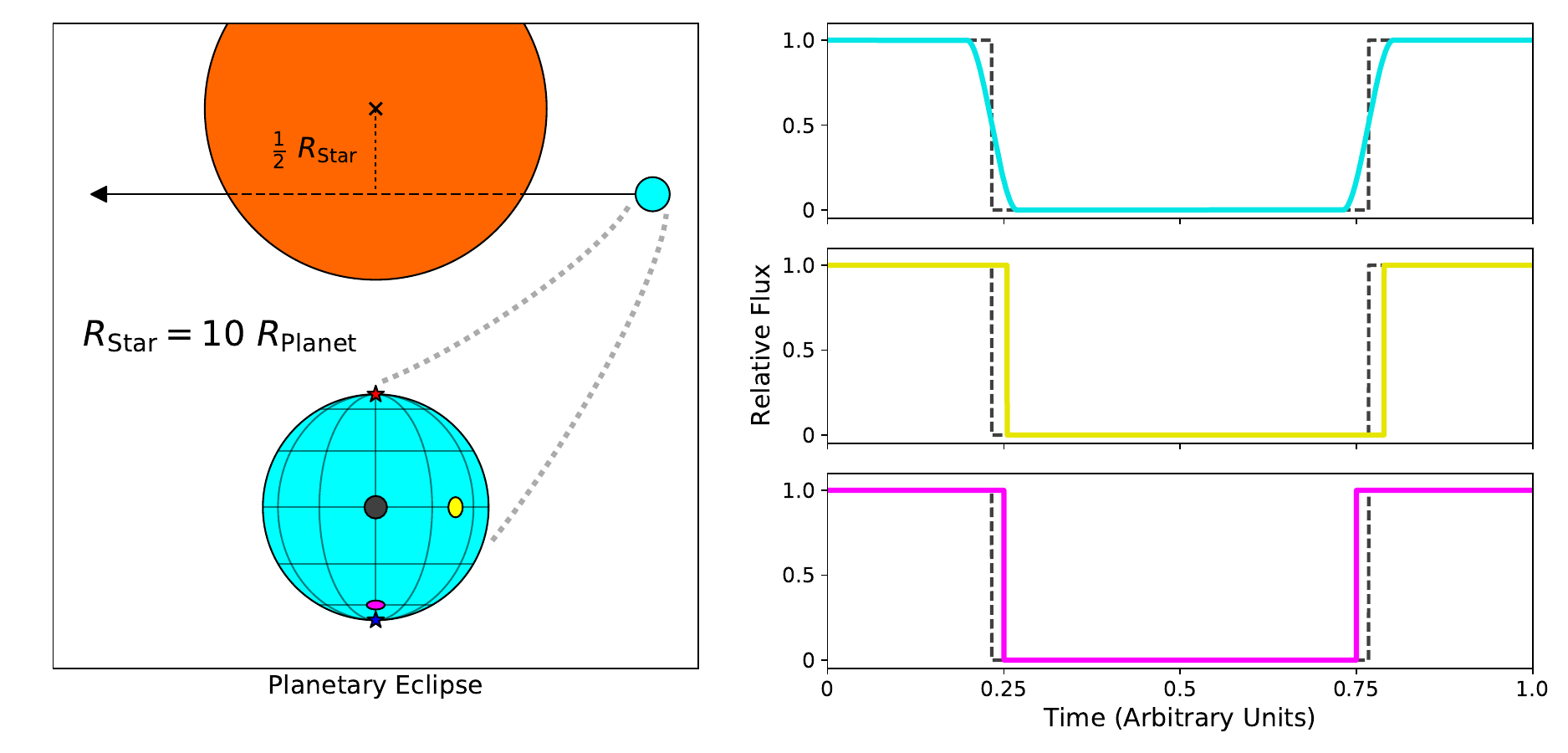}
\caption{Schematic showing the dominant signals present in secondary eclipse lightcurves.  The teal lightcurve denotes a uniform planet \citep[``occultuniform'' of][]{Mandel_Agol_2002}.  The gray dashed line is the eclipse shape if the planetary flux is concentrated in the center of the disk \citep[a caricature of limb-darkening or the dayside hot-spot of a short-period planet;][]{Rauscher_2007}. 
The yellow and magenta lines show the lightcurves for the same concentrated bright spot at different locations on the planetary disk. 
In particular, the yellow line shows the effect of a longitudinal offset predicted by \cite{Williams_2006} and measured by \cite{Agol_2010}. 
The magenta line shows the effect of a latitudinal offset predicted by \cite{Rauscher_2007} and measured by \cite{Majeau_2012} and \cite{deWit_2012}.
All of these signals suffer degeneracies with system geometry, notably impact parameter, planet/star radius ratio, and orbital eccentricity; ancillary constraints are therefore crucial to eclipse mapping \citep{deWit_2012,Rauscher_2018}.  Moreover, some degeneracies can be lifted with spectrally resolved data \citep{Dobbs-Dixon_2015}: the planetary map may be wavelength-dependent, but the system geometry cannot. Figure from J.C.~Schwartz  (private communication).}
\label{fig:delta_map_eclipses}
\end{center}
\end{figure}

If a planet passes directly behind its star, then one may map its day side. \hbindex{Eclipse mapping} can be applied to either planetary emission or reflected light, but the contrast ratio makes the latter daunting.  
While the eclipse depth is proportional to the integrated day side brightness of the planet, the detailed shape of ingress and egress is a function of the spatial distribution of flux on the planet's day side (Figure~\ref{fig:delta_map_eclipses}). Although eclipses of planets by their host star have so far garnered the most attention, occultations by moons or other planets are also possible, especially in packed planetary systems like TRAPPIST-1 \citep[][]{Veras_2017, Luger_2017}.

Fast codes are now available to produce occultation lightcurves numerically \citep[SPIDERMAN;][]{Louden_2018} and analytically \citep[starry;][]{Luger_2019a}. The sharp kernel edge during an occultation---the stellar limb---increases the spatial resolution achievable in eclipse maps, and a larger impact parameter provides better latitudinal resolution  \citep{2023arXiv231014245B}. 

\section{Current Results \& Future Prospects}
The atmospheres of a few dozen exoplanets have been mapped via their thermal phase curves, e.g., the  \href{https://research.iac.es/proyecto/exoatmospheres/table.php}{ExoAtmospheres} and \href{https://exoplanet.eu/home/}{exoplanet.edu} databases. These phase measurements has so far been mostly IRAC photometry from the \hbindex{Spitzer Space Telescope} \citep[e.g.,][]{2021MNRAS.504.3316B,2022AJ....163..256M}, or spectroscopy with the Hubble Space Telescope's WFC3  \citep[e.g.,][]{Stevenson_2014,2018AJ....156...17K}. We send the interested reader to the brief history of thermal phase curves in \cite{2022ExA....53..417C}. Phase measurements from the \hbindex{James Webb Space Telescope} have started being published, but so far focusing primarily on photometry  \citep{2023ApJ...943L..17M,2023arXiv230106350B,2023Natur.620...67K}. Rotational phase variations of brown dwarfs have also been measured and converted into brightness maps \citep[e.g.,][and chapter by Artigau]{Apai_2017}. 

The \hbindex{Ariel} mission could measure spectroscopic phase curves for dozens---if not hundreds---of planets and brown dwarfs \citep{2022ExA....53..417C}. The wavelength coverage of \emph{Ariel} and JWST should be able to discriminate between thermal emission and reflected light \citep{Schwartz_Cowan_2015,Parmentier_2016,Keating_2017}, as well as disentangling phase variations, star spots, Doppler beaming, ellipsoidal variations, and gas accretion \citep{Cowan_2012, Knutson_2012, Esteves_2015,Dang_2018,Bell_2019}.    

\begin{figure}
\begin{center}
\includegraphics[scale=0.25]{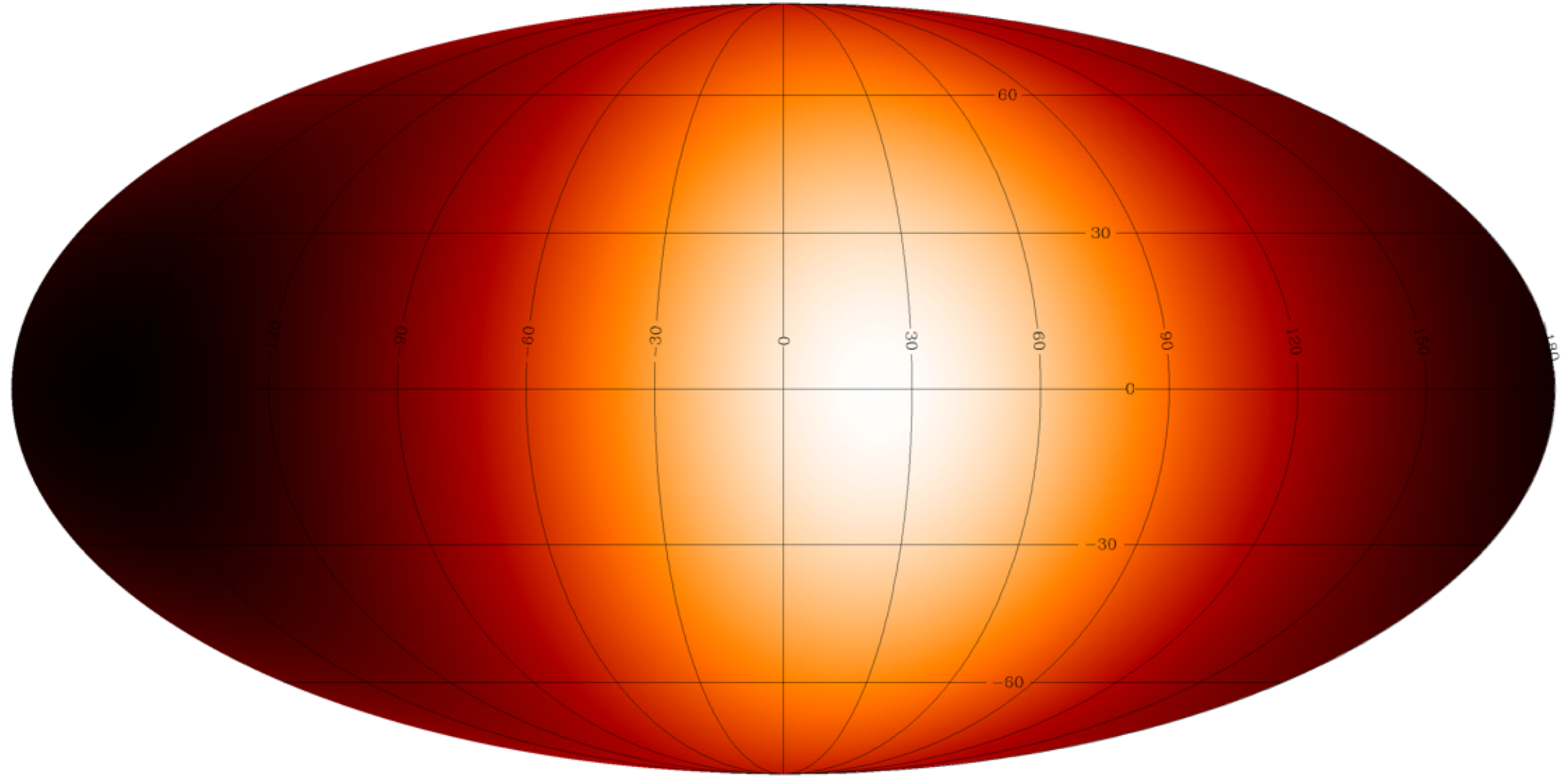}
\caption{Two-dimensional thermal map of the hot Jupiter HD~189733b, based on mid-infrared phase and eclipse measurements from the Spitzer Space Telescope \citep[from][]{Majeau_2012}. The sub-stellar point is in the center of the map. The equatorial hotspot indicates a small obliquity, while the eastward offset is probably due to super-rotating zonal winds \citep{Showman_2002}. 
}
\label{fig:HD189_map}
\end{center}
\end{figure}

Eclipse mapping of hot Jupiters has so far been performed with \emph{Spitzer}/IRAC photometry \citep{Majeau_2012, deWit_2012} and JWST/NIRISS spectroscopy \citep{2023Natur.620..292C}.  Figure~\ref{fig:HD189_map} shows the combined phase+eclipse map of HD 189733b from \cite{Majeau_2012} showing both longitudinal and latitudinal information, making this a coarse 2D map. The high signal-to-noise and spectral resolution of JWST should enable 3D temperature maps of the daysides of hot Jupiters, and 2D (longitude and altitude) temperature maps of their nightsides. Eclipse mapping of longer-period planets, which may not be tidally locked, could reveal their spin axis  \citep{2017ApJ...846...69R,2023AJ....165...24A, 2023AJ....165..261R}. 

\begin{figure}
\begin{center}
\includegraphics[scale=0.7]{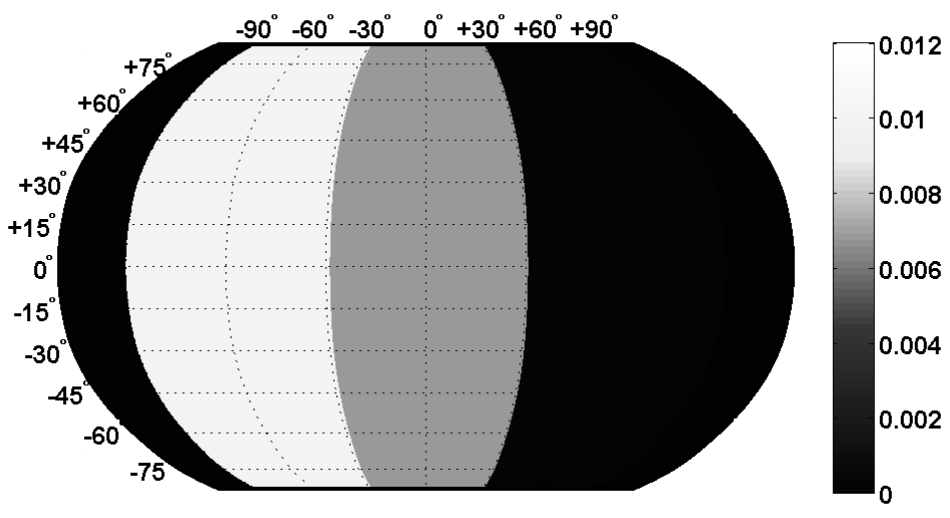}
\caption{Map of normalized intensity, $I_p/I_*$, for the planet Kepler-7b based on orbital phase variations measured by the Kepler mission \citep[adapted from][]{Demory_2013}. The intensity is interpreted as reflected light, and hence is proportional to albedo, $A_g = (I_p/I_*)(a/R_*)^2$, and the bright region on the planet's western hemisphere has an albedo of 0.64, suggestive of reflective clouds \citep[][]{Sudarsky_2000, Roman_2017}.  A possible explanation for the cloud pattern is that particles condense on the planet's cooler nightside, are advected to the dayside by eastward winds, and sublimate when the star is overhead \citep[][]{Heng_Demory_2013, Parmentier_2016}.}
\label{fig:Kepler7_map}
\end{center}
\end{figure}

Ground-based direct imaging could detect variations in thermal emission from directly-imaged planets due to rotation of patchy clouds in and out of view \citep{2021MNRAS.503..743B}, as is currently performed for brown dwarfs. The next generation of ground based telescopes should enable both photometric mapping and Doppler tomography of giant planets \citep{Crossfield_2014}, while the \hbindex{Large Interferometer for Exoplanets}  \citep{2022A&A...664A..21Q} could map the climates of temperate terrestrial planets \citep[e.g.,][]{Gaidos_Williams_2004,2012ApJ...757...80C,2023ApJ...946...82M}.  

\begin{figure}
\begin{center}
\includegraphics[scale=0.40]{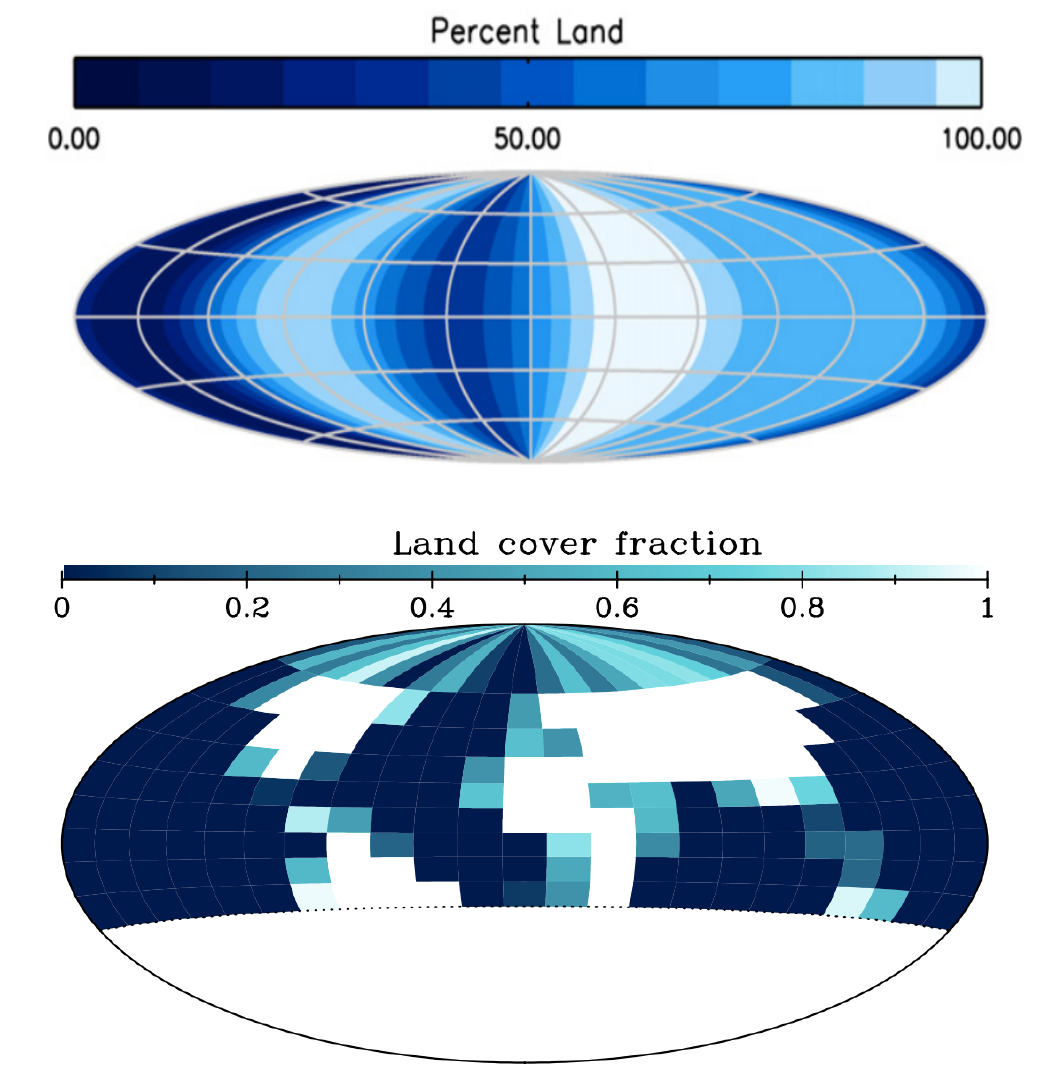}
\caption{\emph{Top:} A one-dimensional color map of Earth based on 24 hours of disk-integrated photometry obtained by the Deep Impact spacecraft as part of the EPOXI mission \citep[from][]{Cowan_2009}.  \emph{Bottom:} A two-dimensional surface map recovered from simulated full-orbit multi-band observations of a cloud-free Earth twin \citep[from][]{Kawahara_2010}; in this simulation the observer was at northern latitudes and hence could not map the southern portion of the planet. One can readily identify the major landforms and oceans in either of these maps, unambiguously identifying this planet as habitable. Most reflected light mapping studies have validated their methods using Earth-as-an-exoplanet \citep[][]{2020plas.book..379R}.}
\label{fig:Fujii_map}
\end{center}
\end{figure}

\cite{Demory_2013} constructed an albedo map of the hot Jupiter Kepler-7b based on phase curves from the \hbindex{Kepler} mission (Fig.~\ref{fig:Kepler7_map}). NASA's \hbindex{Grace Roman Space Telescope} will directly image Jupiter analogs in reflected light, hence enabling rotational mapping of their clouds.  The \hbindex{Habitable Worlds Observatory} will enable reflected light surface mapping and spin determination for terrestrial planets \citep[Figure~\ref{fig:Fujii_map};][]{Ford_2001,Palle_2008, Oakley_2009, Cowan_2009, Cowan_2011, Kawahara_2010, Kawahara_2011, Fujii_2012, Cowan_Strait_2013, Schwartz_2016, Kawahara_2016, Fujii_2017, Jiang_2018, Farr_2018, Luger_2019, Fan_2019, 2020ApJ...894...58K, Aizawa_2020, Nakagawa_2020}.

Surface water is the definition of planetary habitability \citep{Kasting_1993}, while the bimodal surface character of Earth---large oceans and exposed continents---may be crucial to its long-term \hbindex{habitability} \citep{1981JGR....86.9776W,Abbot_2012, Foley_2015,Foley_2019,Dorn_2018, Foley_Smye_2018,Honing_2019}. Constructing surface maps of terrestrial exoplanets will therefore be a step towards understanding habitable environments outside of the Solar System and indeed to finding inhabited worlds \citep[for a review see][]{2019asbi.book..441S}.

\hbindex{Clouds} are a blessing and a curse to exo-cartography. From Earth to brown dwarfs, clouds contribute to the spatial inhomogeneity that make exo-cartography interesting and feasible. But clouds also mask underlying features and often vary in time.  Ongoing efforts are demonstrating how to map clouds and devising schemes to minimize their effects \citep{Cowan_2009,Cowan_Strait_2013,Fujii_2017,Jiang_2018} in order to catch glimpses of the planetary surface below \citep{2020ApJ...900...48K,2022MNRAS.511..440T}.

\begin{acknowledgement}
The authors are grateful to International Space Science Institute for hosting the Exo-Cartography workshops (2016--2017) during which the first version of this review was drafted.  N.B.~Cowan acknowledges the financial support and stimulating environment of the Trottier Space Institute and l'Institut de recherche sur les exoplan\`etes.  
The authors thank J.C.~Schwartz for creating many of the figures, as well as T.~Bell, D.~Keating, E.~Rauscher, and T.~Robinson for providing useful feedback on an earlier draft of this review.   
\end{acknowledgement}

\section*{Cross-References}
\begin{itemize}
\item ``Detecting Habitability''
\item ``Characterization of Exoplanets: Secondary Eclipses''
\item ``Characterization of Exoplanets: Observations and Modeling of Orbital Phase Curves''
\item ``Variability of Brown Dwarfs'' 
\item ``Spectroscopic Direct Detection of Exoplanets''
\end{itemize}

\bibliographystyle{spbasicHBexo}  
\bibliography{Mapping_Review} 
\end{document}